\documentclass{article}

\usepackage{microtype}
\usepackage{graphicx}
\usepackage{subcaption}   %
\usepackage{booktabs}

\usepackage{hyperref}

\usepackage[preprint]{icml2026}

\usepackage{amsmath}
\usepackage{amssymb}
\usepackage{mathtools}
\usepackage{amsthm}

\usepackage[capitalize,noabbrev]{cleveref}

\theoremstyle{plain}
\newtheorem{theorem}{Theorem}[section]

\theoremstyle{definition}
\newtheorem{definition}[theorem]{Definition}

\theoremstyle{remark}

\usepackage{pifont}
\usepackage{tcolorbox}
\tcbuselibrary{breakable}
\usepackage{multirow}
\usepackage{caption}

\newcommand{\finding}[2]{
\begin{center}
\begin{tcolorbox}[leftrule=0mm,toprule=0mm,bottomrule=0mm,rightrule=0mm,
                  left=1pt,right=2pt,top=0pt,bottom=0pt,breakable]
\textbf{Answer to RQ{#1}:}
{#2}
\end{tcolorbox}
\end{center}
}

\icmltitlerunning{Benchmarking Coding Agents on Project-Level Test Evolution}

\begin{document}

\twocolumn[
\icmltitle{Breaking, Stale, or Missing? Benchmarking Coding Agents \\
           on Project-Level Test Evolution}

\icmlsetsymbol{equal}{*}

\begin{icmlauthorlist}
\icmlauthor{Ye Shang}{nju}
\icmlauthor{Quanjun Zhang}{njust}
\icmlauthor{Haichuan Hu}{njust}
\icmlauthor{Chunrong Fang}{nju}
\icmlauthor{Liang Xiao}{njust}
\icmlauthor{Zhenyu Chen}{nju}
\end{icmlauthorlist}

\icmlaffiliation{nju}{State Key Laboratory for Novel Software Technology, Nanjing University, Nanjing, China}
\icmlaffiliation{njust}{Nanjing University of Science and Technology, Nanjing, China}

\icmlcorrespondingauthor{Zhenyu Chen}{zychen@nju.edu.cn}

\icmlkeywords{test evolution, software testing, benchmark, AI agents, LLM agents}

\vskip 0.3in
]

\printAffiliationsAndNotice{}

\begin{abstract}
As production code evolves, the associated test suite must co-evolve to remain effective. Existing benchmarks for test evolution operate at method-level granularity with pre-paired inputs, bypassing the critical task of locating affected tests from the full project and excluding the need for new tests entirely. We present \textbf{TEBench}, the first project-level benchmark for test evolution. Given a project repository and a code-changing commit, TEBench requires systems to autonomously identify tests requiring modification, determine where new tests are needed, and produce the corresponding test patch. We construct TEBench through a four-stage filtering pipeline over projects from the Defects4J ecosystem, curating 314 task instances from 10 projects with developer-written ground truth. Each instance is annotated with one or more of three evolution types: Test-Breaking (tests that fail), Test-Stale (tests that pass but no longer meaningfully validate the updated behavior), and Test-Missing (new tests needed for introduced behavior). We evaluate seven configurations spanning three industrial agent frameworks (Claude Code, Codex CLI, and OpenCode) and six base models, alongside a heuristic baseline. All seven configurations converge on an identification F1 of 45.7\% to 49.4\%, revealing a shared performance ceiling that holds across both agent frameworks and base models. Test-Stale is the most challenging type, with an average F1 of approximately 36\%, since configurations rely on execution failure signals and lack proactive semantic reasoning. On the update task, configurations produce highly executable test modifications whose surface form nonetheless diverges substantially from developer-written ground truth. Analysis of execution trajectories reveals a reactive ``execute-fail-fix'' loop that succeeds for breaking tests but structurally cannot address stale or missing tests.
TEBench is publicly available at \url{https://github.com/iSEngLab/TEBench}, with a continuously updated leaderboard at \url{https://tebench-leadership.vercel.app}.
\end{abstract}

Software systems evolve continuously, with production code undergoing frequent modifications to fix bugs, add features, and refactor implementations. As production code changes, the associated test suite should co-evolve to remain effective. This challenge, known as \textit{test evolution}, is pervasive in practice, yet developers often struggle to systematically identify all tests affected by a given change across a project. Some tests begin to fail due to changed interfaces or updated output formats; others continue to pass but silently lose their ability to validate the behavior they were designed to check; still others are simply absent, as newly introduced functionality lacks any corresponding test. Left unaddressed, these issues lead to gradual test suite degradation that silently undermines software quality.

A growing body of research has addressed the problem of test evolution~\cite{hu2023identify, chi2025reaccept, sun2023revisiting, zhang2025unit}. Due to the limited context window and reasoning capability of earlier techniques such as fine-tuned CodeT5 models~\cite{hu2023identify}, existing approaches adopt a method-level input formulation $\langle m, m', t \rangle$ that pairs the original and updated production methods with an associated test method. This design presupposes that an obsolete test $t$ has already been selected, structurally bypassing the identification step and restricting the problem scope to two categories of \textit{obsolete tests}: tests that fail after the code change, and tests that still pass but whose coverage of the changed code has degraded. The possibility of missing tests, where new behavior lacks any corresponding test, is excluded entirely. With the rapid advancement of large language models and autonomous coding agents, benchmarks should evolve to better reflect real-world development scenarios. This paradigm shift has already occurred in adjacent fields: SWE-bench~\cite{jimenez2024swe} and SWT-bench~\cite{muendler2024swt} elevated the task from patching isolated functions and generating method-level tests (as in Defects4J~\cite{just2014defects4j}) to resolving GitHub issues and reproducing bugs across entire repositories, catalyzing rapid progress in coding agent development. To bridge this gap in test evolution, we lift the input from method-level to project-level and refine the problem taxonomy accordingly. We subdivide obsolete tests into \textit{Test-Breaking}, where the test fails after the change, and \textit{Test-Stale}, where the test still passes but no longer meaningfully validates the updated behavior. We further introduce \textit{Test-Missing} to capture the need for new tests that cover behavior introduced by the change. Together, these three types constitute a more complete characterization of test evolution that aligns with real-world practice.

In this paper, we present \textbf{TEBench} (\textbf{T}est \textbf{E}volution \textbf{Bench}mark), to the best of our knowledge, the first project-level benchmark for test evolution. TEBench defines a new task formulation: given a project repository and a commit that modifies production code, the system must autonomously identify tests requiring modification and determine where new tests are needed across the entire project. We construct TEBench through a four-stage filtering pipeline over projects from the Defects4J~\cite{just2014defects4j} ecosystem, ultimately curating 314 high-quality task instances with developer-written ground truth from 10 real-world open-source Java projects. Each task instance is classified into one or more of the three evolution types defined above, and evaluated through a two-dimensional metric framework that separately measures identification accuracy and update quality.

Using TEBench, we conduct the first systematic evaluation of LLM-based systems on the test evolution task. We evaluate seven configurations spanning three industrial agent frameworks (Claude Code, Codex CLI, and OpenCode) and six base models, alongside a heuristic baseline. All seven configurations converge on an identification F1 of 45.7\% to 49.4\%, with less than four percentage points separating them, and the convergence holds across both agent frameworks and base models, indicating that the bottleneck lies in the inherent task difficulty rather than in any specific configuration. Test-Stale emerges as the most challenging evolution type, with an average F1 of approximately 36\%, since configurations rely almost entirely on execution failure signals and lack the ability to proactively reason about semantic test adequacy. On the update task, configurations produce highly executable test modifications whose surface form nonetheless diverges substantially from developer-written ground truth, indicating that executability is far from a sufficient proxy for update quality. Even exhaustive structural dependency analysis achieves only 66\% Recall, leaving roughly one-third of affected tests undetectable through direct dependency tracing alone.

The main contributions of this paper are as follows:
\begin{itemize}
    \item \textbf{New Dimension.} We propose the \textit{project-level test evolution} task, which requires systems to autonomously identify tests requiring modification and determine where new tests are needed from the full project context, extending beyond the method-level paired formulations of prior work.
    \item \textbf{Benchmark.} We construct \textbf{TEBench}, to the best of our knowledge the first benchmark for project-level test evolution, comprising 314 high-quality task instances with developer-written ground truth from 10 real-world Java projects, covering three evolution types (Test-Breaking, Test-Stale, and Test-Missing), accompanied by a two-dimensional evaluation framework for identification and update quality.
    \item \textbf{Evaluation Study.} We conduct the first systematic evaluation of seven LLM-based configurations spanning three industrial agent frameworks and six base models, revealing a shared performance ceiling, type-specific difficulty patterns, and limitations in proactive semantic reasoning.
\end{itemize}

\section{Motivation and Task Definition}
\label{sec:motivation}
 
\subsection{Motivating Example}
\label{sec:motivation:example}
 
\begin{figure*}[htbp]
  \centering
  \includegraphics[width=\linewidth]{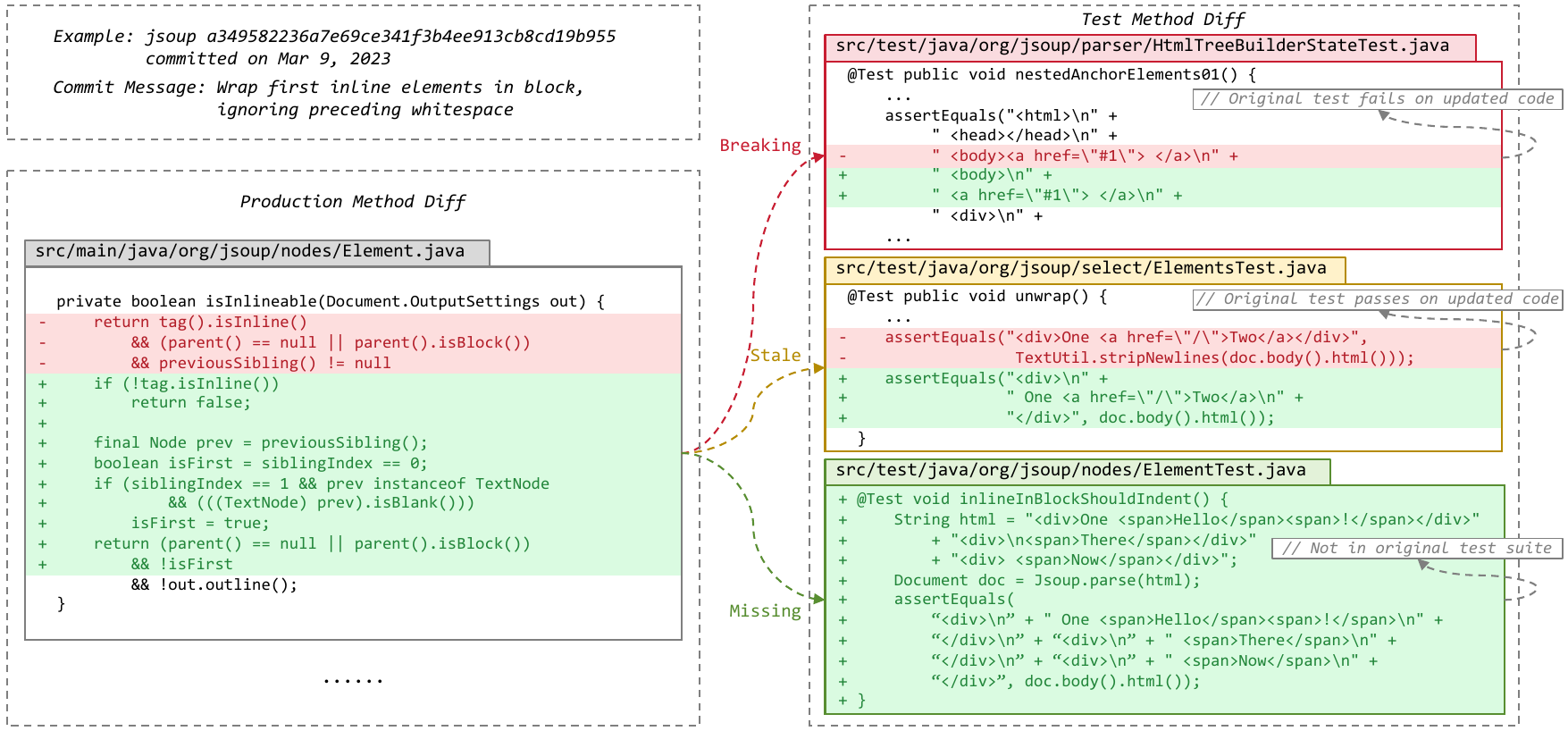}
  \caption{Motivating example from \textsc{jsoup}: a change to \texttt{isInlineable()} in \texttt{Element.java} impacts multiple test files across different packages, exhibiting three types of test evolution.}
  \label{fig:motivation}
\end{figure*}

We illustrate the complexity of real-world test evolution through a concrete example. Figure~\ref{fig:motivation} shows a commit from \textsc{jsoup}, a widely used Java HTML parsing library with over 11K stars on GitHub. This commit is also included as a task instance in TEBench.
 
The commit message states: \textit{``Wrap first inline elements in block, ignoring preceding whitespace.''}
This change fixes the pretty-printing logic in \texttt{isInlineable()} method of \texttt{Element.java} file. In the original implementation, the decision of whether an inline element should be wrapped and indented inside a block element was governed by a simple boolean condition. The revised implementation introduces an additional check for a previously overlooked edge case: when the first inline element in a block is preceded by a blank text node, it should still be treated as the block's first child and therefore be indented, rather than rendered inline.
 
Although the code change is localized to a single production method, its impact on the test suite is both broad and heterogeneous, spanning multiple test files in different packages.
 
\noindent\textbf{Impact 1: Test failure.}
Some existing tests fail immediately after the change. For example, a test such as \texttt{nestedAnchorElements01()} now observes different pretty-printing output: anchor elements that were previously rendered inline are instead wrapped onto indented new lines when preceded by blank text nodes. The same failure pattern also appears in the other two tests across different packages.
 
\noindent\textbf{Impact 2: Silent quality degradation.}
Other tests continue to pass, yet become semantically outdated. In particular, the test \texttt{unwrap()} still succeeds because it applies \texttt{TextUtil.stripNewlines()} before comparing outputs, which masks the formatting change introduced by the updated logic. As a result, the test remains executable but no longer validates the intended pretty-printing behavior. The developer subsequently revised this test to compare directly against the formatted output.
 
\noindent\textbf{Impact 3: Uncovered new behavior.}
The developer also added a new test, \texttt{inlineInBlockShouldIndent()}, to cover scenarios that were previously untested. This test verifies that inline elements inside a block are consistently indented across several representative inputs, filling a gap in the original test suite.
 
This example highlights a key characteristic of real-world test evolution: even a small and localized code change can induce diverse and non-obvious impacts on tests distributed across the project. In practice, a single commit may modify multiple production methods across several files, each affecting a different subset of tests, thereby creating a complex many-to-many relationship between code changes and test impacts. This naturally raises the question: can existing benchmarks capture such complexity?
 
\subsection{Limitations of Existing Benchmarks}
\label{sec:motivation:limitations}
 
\begin{table}[htbp]
\centering
\caption{Comparison of test evolution benchmarks.}
\label{tab:benchmark_comparison}
\renewcommand{\arraystretch}{1.2}
\resizebox{0.5\textwidth}{!}{
\begin{tabular}{lccccc}
\toprule
\textbf{Benchmark} & \textbf{Granularity} & \textbf{Input} & \textbf{Identification} & \textbf{Update} & \textbf{Output} \\
\midrule
SITAR~\cite{wang2021understanding} & Method & $\langle m, m', t \rangle$ & Paired & No & Binary label \\
CHOSEN~\cite{sun2023revisiting} & Method & $\langle m, m', t, t' \rangle$ & Paired & No & Binary label \\
CEPROT~\cite{hu2023identify} & Method & $\langle m, m', t \rangle$ & Paired & B + S & Updated $t'$ \\
REACCEPT~\cite{chi2025reaccept} & Method & $\langle m, m', t \rangle$ & Paired & B + S & Updated $t'$ \\
TaRBench~\cite{yaraghi2025automated} & Method & $\langle P, m, m', t \rangle$ & Assumed & B & Repaired $t'$ \\
Synter~\cite{liu2024fix} & Method & $\langle P, m, m', t \rangle$ & Assumed & B & Repaired $t'$ \\
Tool-Bench~\cite{rahman2025utfix} & Method & $\langle P, m, m', t \rangle$ & Assumed & B + S & Repaired $t'$ \\
Updates4J~\cite{zhang2025unit} & Method & $\langle P, m, m', t \rangle$ & Assumed & B + S & Updated $t'$ \\
\midrule
SWE-bench~\cite{jimenez2024swe} & Project & $\langle P, \textit{Issue} \rangle$ & N/A & N/A & Code patch \\
SWT-bench~\cite{muendler2024swt} & Project & $\langle P, \textit{Issue} \rangle$ & N/A & N/A & Repro. test \\
\midrule
\textbf{TEBench} & \textbf{Project} & $\boldsymbol{\langle P, \Delta\rangle}$ & \textbf{Autonomous} & \textbf{B + S + M} & \textbf{Test patch} \\
\bottomrule
\end{tabular}
}
\end{table}
 
Table~\ref{tab:benchmark_comparison} compares TEBench with existing test evolution benchmarks. In the identification dimension, existing benchmarks either adopt a \textit{Paired} setting, where code--test associations are pre-paired for classification, or an \textit{Assumed} setting, where the affected test is directly given as input. By contrast, TEBench uses an \textit{Autonomous} setting, requiring the system to identify affected tests independently at the project level. In the update dimension, prior benchmarks mainly cover Breaking and, in some cases, Stale tests, while TEBench further includes Missing tests. We identify three limitations that prevent current benchmarks from capturing the complexity illustrated above.
 
\noindent\textbf{Limitation 1: Method-level granularity.}
Existing test evolution benchmarks are formulated at the method level, typically taking as input a tuple such as $\langle m, m', t \rangle$, where $m$ and $m'$ denote the original and updated production method, and $t$ denotes the associated test method. This formulation reduces the task to a bounded and pre-identified one-to-one setting (one production method change to one test method). While such formulations are suitable for studying localized test co-evolution, they abstract away the project-level reasoning required in realistic development settings. In the \textsc{jsoup} example, a change to a single production method affects tests across three different packages, requiring cross-file and cross-module reasoning that method-level benchmarks cannot assess.
 
\noindent\textbf{Limitation 2: Identification is bypassed.}
Because the code–test association is already provided, existing benchmarks structurally bypass the test identification task. As shown in Table~\ref{tab:benchmark_comparison}, prior work either provides pre-paired associations for classification (\textit{Paired}) or directly gives the affected test for repair (\textit{Assumed}). In practice, however, identifying which tests among hundreds of files require attention is the first, and often one of the hardest, steps after a code change.
 
\noindent\textbf{Limitation 3: Incomplete coverage of evolution types.}
Because the input always includes an existing test $t$, the output is inherently restricted to a modified version $t'$, thereby excluding the possibility of generating entirely new tests. Consequently, existing benchmarks cover at most tests that fail after a code change and tests that still pass but nevertheless require updates. They do not account for cases in which new behavior is introduced or exposed, but no corresponding test yet exists. In practice, however, adding new tests in response to a code change is not an independent test generation task; it is contextually grounded in the commit itself, motivated by the need to cover behavior specifically introduced by the commit change. In this sense, such tests addition constitutes a natural form of test evolution rather than general-purpose test generation. By restricting the output to modifications of a given $t$, prior benchmarks artificially narrow the scope of test evolution, excluding a response pattern that developers regularly employ, as illustrated by Impact~3 in the motivating example.
 
\subsection{Task Definition}
\label{sec:motivation:task}
 
To address these limitations, we formulate the task of \textit{project-level test evolution} as follows.
 
\begin{definition}[Project-Level Test Evolution]
\label{def:task}
Given a project repository $P$ and a commit change $\Delta$, the system must:
\begin{enumerate}
    \item \textbf{Identify}: determine which existing tests require modification and whether additional tests should be created;
    \item \textbf{Update}: produce the corresponding test patch, including modifications to obsolete tests and any newly generated tests.
\end{enumerate}
\end{definition}
 
Unlike prior method-level formulations, where the relevant code--test association is given as part of the input, our task requires the system to navigate the codebase autonomously, locate affected tests, and generate appropriate updates or additions. As illustrated by the motivating example, we categorize test evolution instances into three types according to how existing tests behave after the code change and how the developer responds.
 
\noindent\textbf{Test-Breaking.}
An existing test $t \in T$ fails to compile or execute after $\Delta$ is applied, and the developer modifies $t$ in the ground truth (GT) to restore correctness. In the motivating example, this corresponds to Impact~1: several tests fail because their expected output strings no longer match the updated formatting behavior.
 
\noindent\textbf{Test-Stale.}
An existing test $t \in T$ still passes after $\Delta$ is applied, but the developer nonetheless updates $t$ in the GT so that it better reflects the revised semantics of the code. In the motivating example, this corresponds to Impact~2: a test remains executable but no longer meaningfully validates the formatting behavior because its comparison logic masks the change.
 
\noindent\textbf{Test-Missing.}
The developer adds a new test method $t_{\text{new}} \notin T$ in the GT to cover behavior introduced or exposed by $\Delta$. In the motivating example, this corresponds to Impact~3: a new test is added to verify consistent indentation behavior across several representative scenarios.
 
The first two types, Test-Breaking and Test-Stale, correspond to what prior work broadly refers to as \textit{obsolete tests}~\cite{hu2023identify,chi2025reaccept,sun2023revisiting}. We further distinguish them according to whether the test fails on the updated code. The third type, Test-Missing, extends beyond the scope of prior work by capturing the need for entirely new tests.

\section{Benchmark Construction}
\label{sec:benchmark}

\subsection{Task Construction Pipeline}
\label{sec:benchmark:pipeline}

\begin{figure}[htbp]
  \centering
  \includegraphics[width=\linewidth]{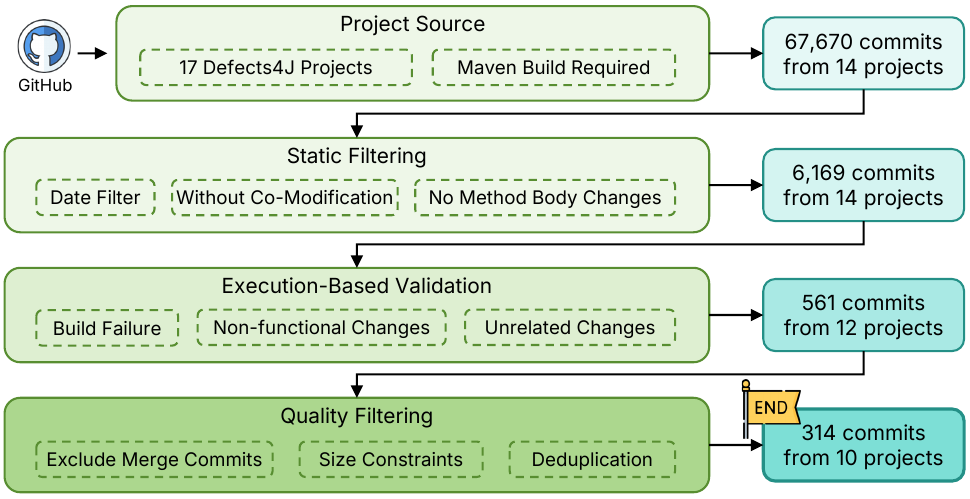}
  \caption{Task construction pipeline. Numbers indicate the remaining commits after each stage.}
  \label{fig:pipeline}
\end{figure}

We designed a multi-stage filtering pipeline to extract high-quality test evolution task instances from real-world open-source Java projects. Figure~\ref{fig:pipeline} illustrates the overall process.

\subsubsection{Project Source.}
TEBench draws its projects from the Defects4J~\cite{just2014defects4j} ecosystem, a widely-used benchmark repository in software engineering research that curates real-world Java projects with high-quality test suites. Starting from 17 Java open-source projects in Defects4J, we excluded 3 projects that do not use Maven as their build system, since our automated test execution and coverage analysis pipeline relies on the Maven Surefire Plugin and JaCoCo. The remaining 14 Maven-based projects span diverse functional domains, including data parsing, text processing, encoding, compression, mathematical computation, chart rendering, and general-purpose language utilities, yielding a total of 67,670 commits.

\subsubsection{Static Filtering}
\label{sec:benchmark:static}
 
We applied a series of static filters to narrow the candidate set without requiring code execution. First, we restricted the time range to commits after January 2019 to ensure Java 8+ syntax compatibility and relevance to modern coding practices. For projects with limited recent history (e.g., commons-math), we relaxed the cutoff to 2016 to maintain a sufficient sample size. Next, we performed file-level scanning to retain only commits that simultaneously modify both production code (under \texttt{src/main/}) and test code (under \texttt{src/test/}), as co-modification is a necessary signal for test evolution. Finally, we used the \texttt{javalang} AST parser to extract method-level change information from each candidate commit, filtering out commits whose modifications are limited to imports, annotations, or comments rather than substantive method body changes. After static filtering, 6,169 commits remained across 14 projects.

\subsubsection{Execution-Based Validation}
\label{sec:benchmark:execution}
 
For each remaining candidate, we constructed isolated execution environments using \texttt{git worktree} to validate the relationship between code changes and test modifications. Specifically, for each commit we built two versions: one with the full commit applied (both production and test changes), and one with only the non-test changes applied while retaining the original test suite. We executed the test suite on both versions and collected line and branch coverage via JaCoCo, enabling us to determine whether the test modifications address actual test failures or contribute to coverage improvements. The detailed version structure is formalized in Section~\ref{sec:benchmark:formulation}.
Based on the execution results, we excluded commits in three categories: (1)~commits where the project fails to compile on the historical version, or where pre-existing test failures unrelated to the commit's code changes are observed; (2)~commits where the test modifications have no measurable impact on test outcomes or code coverage, indicating non-functional changes such as test reorganization, comment edits, or stylistic adjustments, which account for approximately half of the exclusions; and (3)~commits where the test changes lack a verifiable causal relationship with the production code changes. After execution-based validation, 561 commits remained across 12 projects.

\subsubsection{Quality Filtering}
\label{sec:benchmark:quality}
 
We applied final quality controls to ensure each task instance is suitable for benchmarking. Merge commits were excluded as they represent branch integration rather than individual code evolution; the actual evolution occurs in the constituent commits that we already analyze independently. We constrained test change size to 5--200 lines, a range determined through manual inspection of a stratified sample of candidate commits: commits with fewer than 5 lines of test changes consistently involved superficial modifications such as single-assertion tweaks or comment additions that do not constitute meaningful evolution instances, while those exceeding 200 lines typically involved large-scale refactoring that obscures the causal relationship between specific code changes and test updates. Finally, we performed method-level deduplication using \texttt{(project, ClassName.methodName)} as a composite key, retaining only the earliest commit when the same test method appears in multiple commits. This strategy was adopted after manual review revealed that later commits modifying the same test method predominantly represent iterative refinements rather than independent evolution scenarios, which would otherwise introduce redundancy and inflate task counts artificially. After quality filtering, \textbf{314 task instances from 10 projects} constitute the final TEBench dataset. Four projects were excluded as they yielded insufficient valid task instances after the full filtering process.

\subsection{Task Formulation and Evaluation Protocol}
\label{sec:benchmark:formulation}

\subsubsection{Version Design}
\label{sec:benchmark:versions}

\begin{figure}[htbp]
  \centering
  \includegraphics[width=\linewidth]{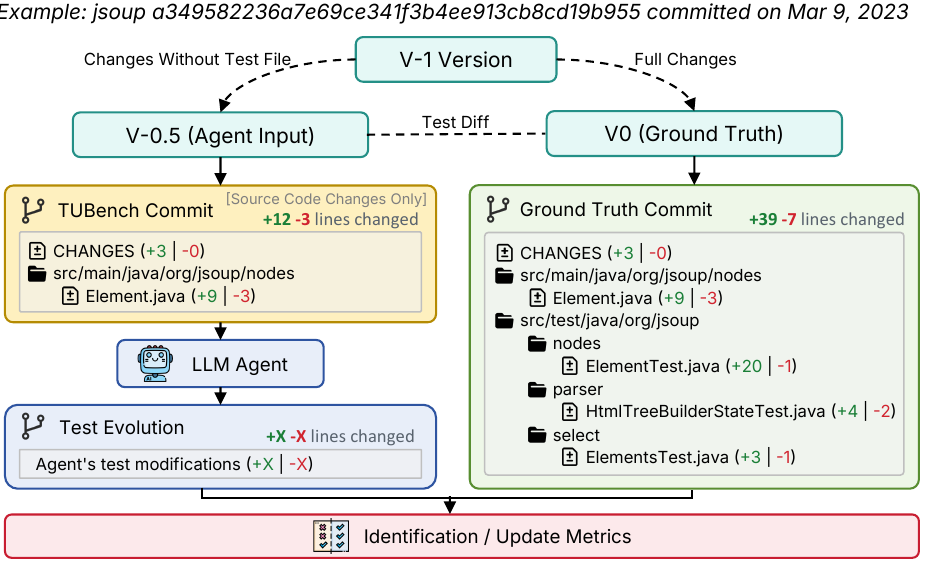}
  \caption{Three-version structure and its dual role in classification and evaluation.}
  \label{fig:version_design}
\end{figure}

Each task instance in TEBench is built around a three-version structure, as illustrated in Figure~\ref{fig:version_design}.
\textbf{$V_{-1}$} represents the project state at the parent commit, before any changes are applied. \textbf{$V_{-0.5}$} is constructed by applying all changes from the commit \textit{except} modifications to test files. This includes production code changes, build configuration updates, and resource file modifications, simulating the real-world scenario where a developer has committed code changes but has not yet updated the corresponding tests. \textbf{$V_0$} represents the full commit state including the developer's actual test modifications, serving as the GT.
This version structure also serves as the basis for classifying task instances into Breaking, Stale, and Missing, as defined in Section~\ref{sec:motivation:task}.
During evaluation, $V_{-0.5}$ is the project state presented to the coding agent. The agent is informed via its prompt that the most recent commit modified the production code without updating the test suite, and is tasked with identifying and updating any affected tests. The agent can access all project-level information, including commit history, commit messages, and code structure, through its standard tooling. The GT for evaluation is the developer's actual test modifications in $V_0$.

\subsubsection{Identification Metrics}
\label{sec:benchmark:id_metrics}
 
The identification stage evaluates whether the agent correctly locates the tests that require attention. We extract the set of affected tests from the GT and compare it against the set of tests actually modified or added by the agent, computing Precision, Recall, and F1-score.
We adopt different granularities for different change types. Modified and deleted test methods are evaluated at \textit{method-level} granularity: a true positive requires the agent to modify or delete the same test method as in the GT. Newly added test methods are evaluated at \textit{file-level} granularity: a true positive requires the agent to add at least one new test method in the same test file where the GT adds new methods. For modified tests, the agent should precisely identify which existing methods need changes; for new tests, it is unreasonable to expect the agent to predict the exact method names or count chosen by the developer, but it should recognize \textit{where} new tests are needed.

\subsubsection{Update Metrics}
\label{sec:benchmark:update_metrics}
 
The goal of test evolution is not merely to produce passing tests or to maximize coverage, but to align the test suite with the intent behind the code change. We therefore design our update metrics around the developer-written GT as a reference for evolution intent, evaluating agents across three dimensions: executability, coverage alignment, and modification similarity.
 
\smallskip
\noindent\textbf{Executability} measures whether the agent's modifications produce valid, runnable tests. We compile and execute the union of test methods modified by the agent and those in the GT. This serves two purposes: executing the agent's modifications verifies whether they introduce compilation or runtime errors, while executing GT methods that the agent did not modify reveals whether the agent missed broken tests that require repair. We assign a three-level score:
 
\begin{equation}
s_{exec} = \begin{cases} 0,   & \text{if compilation fails} \\ 0.5, & \text{if compilation succeeds but tests fail} \\ 1,   & \text{if all tests pass} \end{cases}
\end{equation}
 
\noindent\textbf{Coverage Overlap} measures how well the agent's tests align with the developer's testing intent for the changed code. We execute both the agent's and the GT test suites, and collect line coverage and branch coverage restricted to the production methods modified by $\Delta$. Rather than measuring absolute coverage improvement, we compute the overlap between the agent's coverage and the GT coverage:
 
\begin{equation}
s_{line} = \frac{|C^{agent}_{line} \cap C^{gt}_{line}|}{|C^{gt}_{line}|}, \quad
s_{branch} = \frac{|C^{agent}_{branch} \cap C^{gt}_{branch}|}{|C^{gt}_{branch}|}
\end{equation}
 
\noindent where $C^{agent}$ and $C^{gt}$ denote the sets of lines or branches covered by the agent's and GT tests, respectively. This design reflects the fact that the goal of test evolution is not to maximize coverage indiscriminately, but to ensure the test suite evolves in alignment with the developer's intent regarding the specific code change.
 
\smallskip
\noindent\textbf{Modification Similarity} measures how closely the agent's test changes resemble the GT, capturing whether the agent makes precise, targeted modifications rather than excessive rewrites. We compute the token-level Jaccard similarity between the agent's and GT test modifications:
 
\begin{equation}
s_{mod} = \frac{|tokens_{agent} \cap tokens_{gt}|}{|tokens_{agent} \cup tokens_{gt}|}
\end{equation}
 
\smallskip
\noindent\textbf{Composite Score.} We combine the three dimensions into a single update score. Executability serves as a gate: if the agent's modifications do not compile, the entire score is zero. When the GT produces no coverage change over the original tests (e.g., the change only updates assertion values), the modification similarity receives full weight:
 
\begin{equation}
s_{update} = s_{exec} \times \begin{cases}
\begin{aligned}[t]
& 0.3\, s_{line} + 0.3\, s_{branch} \\
& \quad + 0.4\, s_{mod},
\end{aligned} & \text{if } |C^{gt}| > 0 \\[6pt]
s_{mod}, & \text{if } |C^{gt}| = 0
\end{cases}
\end{equation}

\subsection{Dataset Statistics}
\label{sec:benchmark:statistics}
 
Table~\ref{tab:project_overview} provides an overview of the 10 projects and 314 task instances in TEBench. Source lines of code (Src LOC) count production code only, excluding test files. B, S, and M denote Breaking, Stale, and Missing, respectively.
 
\begin{table}[htbp]
\centering
\caption{Overview of TEBench.}
\label{tab:project_overview}
\renewcommand{\arraystretch}{1.15}
\resizebox{\columnwidth}{!}{
\begin{tabular}{lrrrrrrr}
\toprule
\textbf{Project} & \textbf{Tasks} & \textbf{Src LOC} & \textbf{Test Files} & \textbf{B} & \textbf{S} & \textbf{M} \\
\midrule
commons-cli         &  18 &    9,716 &   51 &   8 &  12 &   9 \\
commons-codec       &  19 &   25,102 &   84 &  12 &  11 &  12 \\
commons-collections &  23 &   80,241 &  300 &  10 &  14 &  15 \\
commons-compress    &  86 &   92,057 &  260 &  34 &  58 &  53 \\
commons-csv         &  31 &    6,295 &   43 &  22 &  18 &  16 \\
commons-lang        &  69 &  101,573 &  275 &  28 &  46 &  40 \\
commons-math        &   8 &  142,903 &  403 &   8 &   3 &   2 \\
gson                &   1 &   22,329 &  139 &   1 &   0 &   1 \\
jfreechart          &   3 &  211,097 &  361 &   1 &   3 &   1 \\
jsoup               &  56 &   27,390 &   84 &  48 &  42 &  50 \\
\midrule
\textbf{Total}      & \textbf{314} & \textbf{718,703} & \textbf{2,000} & \textbf{172} & \textbf{207} & \textbf{199} \\
\bottomrule
\end{tabular}
}
\end{table}
 
\noindent\textbf{Label Distribution.}
The three evolution types are well-represented across the dataset: Breaking appears in 172 tasks (54.8\%), Stale in 207 (65.9\%), and Missing in 199 (63.4\%). Notably, 219 tasks (69.7\%) carry multiple labels, and 45 tasks (14.3\%) exhibit all three types simultaneously. The most frequent combination is Stale + Missing (105 tasks, 33.4\%), suggesting that when developers recognize quality degradation in existing tests, they often supplement new tests in the same commit. Only 95 tasks (30.3\%) involve a single evolution type, confirming that real-world test evolution is predominantly multi-faceted.
 
\noindent\textbf{Task Complexity.}
Table~\ref{tab:difficulty} summarizes the complexity characteristics of the task instances. The median task involves 4 changed files, 34 lines of source code changes, and 32 lines of test changes, indicating moderate complexity that is challenging yet tractable for current coding agents. The distribution exhibits a long tail: the most complex task spans 20 files with 732 lines of source changes.
 
\begin{table}[htbp]
\centering
\caption{Task complexity statistics.}
\label{tab:difficulty}
\renewcommand{\arraystretch}{1.15}
\resizebox{0.9\columnwidth}{!}{
\begin{tabular}{lrrrrr}
\toprule
\textbf{Metric} & \textbf{Mean} & \textbf{P25} & \textbf{Median} & \textbf{P75} & \textbf{Max} \\
\midrule
Source lines changed    & 74.8 & 13 & 34 & 78 & 732 \\
Test lines changed      & 48.6 & 14 & 32 & 72 & 200 \\
Total files changed     &  4.9 &  3 &  4 &  6 &  20 \\
Test files per task     &  1.8 &  1 &  1 &  2 &  10 \\
Test methods per task   &  4.6 &  2 &  3 &  5 &  74 \\
\bottomrule
\end{tabular}
}
\end{table}
 
\noindent\textbf{Project-Level Characteristics.}
A key motivation for TEBench is that test evolution requires project-level reasoning. Our statistics confirm this: 114 tasks (36.4\%) involve modifications to more than one test file, 63 tasks (20.1\%) span multiple test packages, and 236 tasks (75.2\%) require changes to more than one test method. These numbers demonstrate that a substantial portion of test evolution tasks cannot be adequately captured by method-level benchmarks that assume a one-to-one mapping between code changes and test modifications.
 
\noindent\textbf{Temporal Distribution.}
TEBench spans commits from 2016 to 2025. The majority of tasks (77.4\%) originate from 2020 or later, with 2024--2025 contributing 125 tasks (39.8\%), ensuring that the dataset reflects contemporary development practices and coding conventions.

\section{Experimental Setup}
\label{sec:experiment}

\subsection{Evaluated Systems}
\label{sec:experiment:systems}

We evaluate eight systems organized along two axes: a heuristic baseline and seven LLM-based configurations spanning three coding agent frameworks and six base models. Table~\ref{tab:systems} summarizes all evaluated systems.

\begin{table}[htbp]
\centering
\caption{Evaluated systems.}
\label{tab:systems}
\resizebox{0.85\columnwidth}{!}{
\begin{tabular}{lll}
\toprule
\textbf{Agent Framework} & \textbf{Base Model} & \textbf{Version} \\
\midrule
Heuristic Baseline & ---                & --- \\
\midrule
Claude Code     & Claude Sonnet 4.6  & v2.1.45 \\
Codex CLI       & ChatGPT 5.3 Codex  & v0.114.0 \\
OpenCode        & Claude Sonnet 4.6  & v1.2.16 \\
\midrule
OpenCode        & Qwen3.5            & v1.2.16 \\
OpenCode        & GLM-5              & v1.2.16 \\
OpenCode        & Kimi-K2.5          & v1.2.16 \\
OpenCode        & DeepSeek-V3.2      & v1.2.16 \\
\bottomrule
\end{tabular}
}
\end{table}

\subsubsection{Heuristic Baseline.}
To establish a lower bound on what structural analysis alone can achieve, we implement a static dependency baseline that operates in three steps. First, it extracts changed classes and methods from the source code diff using AST-level analysis via \texttt{javalang}. Second, it scans all \texttt{@Test}-annotated methods in the project, retaining those whose enclosing file imports a changed class and whose method body invokes a changed symbol. Third, it validates candidates by executing them with Maven's Surefire plugin, filtering out methods that cannot be located at runtime. This baseline is designed exclusively for the identification subtask and does not perform test updates, serving as a reference for evaluating how far structural analysis alone can reach in locating affected tests.

\subsubsection{Coding Agents and Base Models.}
We evaluate three widely-adopted industrial coding agent frameworks: Claude Code~\cite{anthropic2025claude} (Anthropic, closed-source), Codex CLI~\cite{openai2025codex} (OpenAI, closed-source), and OpenCode~\cite{opencodecontributors2025opencode} (open-source). These frameworks are paired with six base models that span closed-source flagships and open-weight families: Claude Sonnet 4.6, ChatGPT 5.3 Codex, Qwen3.5~\cite{qwen}, GLM-5~\cite{glm}, Kimi-K2.5~\cite{kimi}, and DeepSeek-V3.2~\cite{deepseek}. Claude Code and Codex CLI are evaluated under their respective default backbones (Claude Sonnet 4.6 and ChatGPT 5.3 Codex). OpenCode is evaluated with five backbones: Claude Sonnet 4.6, Qwen3.5, GLM-5, Kimi-K2.5, and DeepSeek-V3.2, yielding five distinct configurations under a single framework.

All configurations run with default agent settings and are given full access to the project workspace within an isolated environment (Section~\ref{sec:experiment:env}). We adopt a \emph{natural-run} evaluation mode: each configuration receives the unified task prompt and is allowed to freely explore the project, execute tests, inspect coverage reports, and iteratively refine its modifications without any artificial constraints on its problem-solving strategy. Upon completion, we extract the actual modifications produced by each configuration to infer its identification decisions. Specifically, test methods that are modified or deleted are treated as identifications of obsolete tests, while newly added test methods are treated as identifications of missing tests.

\subsection{Task Prompt}
\label{sec:experiment:prompt}

\begin{figure}[htbp]
\centering
\includegraphics[width=\linewidth]{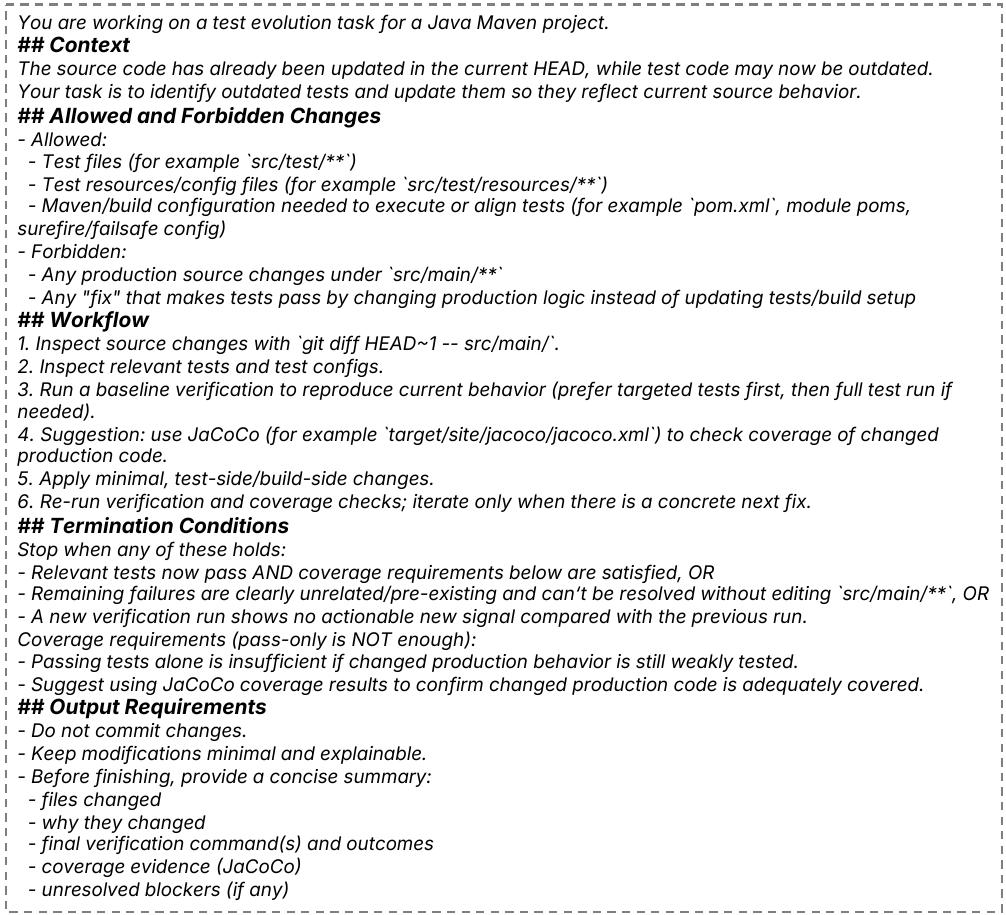}
\caption{Unified task prompt provided to all configurations.}
\label{fig:prompt}
\end{figure}

All configurations receive an identical, commit-type-agnostic task prompt regardless of whether a task instance involves Test-Breaking, Test-Stale, or Test-Missing changes, ensuring that each configuration must independently determine the nature and extent of required updates. The full prompt is presented in Figure~\ref{fig:prompt}.

The prompt is structured around five components that together define the task boundaries. The \emph{Context} component informs each configuration that source code has already been updated while tests may be outdated. The \emph{Allowed and Forbidden Changes} component restricts modifications to test files, test resources, and build configuration, while strictly prohibiting any changes to production source code under \texttt{src/main/}, which mirrors the real-world constraint that the source change has already been committed and the task is solely to bring tests in line. Each configuration is also restricted from inspecting any commits beyond the current HEAD, which prevents it from accessing ground-truth test modifications in subsequent commits. The \emph{Workflow} component suggests a recommended sequence of inspection, verification, and iteration steps, including the use of JaCoCo coverage analysis to assess whether changed production code is adequately exercised, since passing tests alone may mask insufficient coverage of newly introduced behavior. The \emph{Termination Conditions} component defines explicit stopping criteria to prevent configurations from entering infinite repair loops, allowing them to stop when tests pass with adequate coverage, when remaining failures are clearly unrelated to the current commit, or when successive verification runs yield no additional actionable signal. Finally, the \emph{Output Requirements} component instructs configurations to keep modifications minimal and explainable, and to provide a concise summary of changed files, rationale, verification outcomes, and coverage evidence before finishing, which facilitates subsequent automated evaluation.

\subsection{Execution Environment}
\label{sec:experiment:env}

For each task instance, we construct an isolated execution environment based on the $V_{-0.5}$ version defined in Section~\ref{sec:benchmark:formulation}. Specifically, we use Git's \emph{worktree} mechanism to create a dedicated working directory for each task: a new branch is created at the $V_{-0.5}$ state, which contains the updated source code with original tests, and a separate worktree is attached to this branch. This approach provides full filesystem isolation between tasks, as each configuration operates in its own independent copy of the project at the correct historical state, free from interference by other versions or concurrent executions.

Compared to provisioning a separate Docker container per project, the worktree-based approach is significantly more lightweight while achieving equivalent isolation for our purposes, since each task has its own independent source tree, build artifacts, and test execution context while sharing only the read-only Git object store with the main repository. To support reproducibility, we provide a Docker image that bundles all project environments and evaluation scripts in our replication package, allowing other researchers to replicate our experiments without configuring individual project dependencies.

\section{Results and Analysis}
\label{sec:results}

We organize our evaluation around four research questions:

\begin{itemize}
    \item \textbf{RQ1 (Identification):} How effectively can current configurations identify obsolete tests and missing tests in evolving projects?
    \item \textbf{RQ2 (Update):} How effectively can current configurations update obsolete tests and generate missing tests?
    \item \textbf{RQ3 (Task Characteristics):} How do task characteristics influence configuration performance?
    \item \textbf{RQ4 (Failure Analysis):} What are the typical failure modes of configurations on test evolution tasks?
\end{itemize}

\subsection{RQ1: Identification Effectiveness}
\label{sec:results:rq1}

Table~\ref{tab:rq1} presents the identification results across all 314 task instances, reported both overall and per evolution type.

\begin{table}[htbp]
\centering
\caption{Identification results (Precision / Recall / F1, \%). Best F1 per type column among the seven LLM-based configurations is \textbf{bolded}. The heuristic baseline participates only in identification and is reported separately for reference.}
\label{tab:rq1}
\renewcommand{\arraystretch}{1.15}
\resizebox{\linewidth}{!}{
\begin{tabular}{l rrr rrr rrr rrr}
\toprule
& \multicolumn{3}{c}{\textbf{Overall}} & \multicolumn{3}{c}{\textbf{Breaking}} & \multicolumn{3}{c}{\textbf{Stale}} & \multicolumn{3}{c}{\textbf{Missing}} \\
\cmidrule(lr){2-4} \cmidrule(lr){5-7} \cmidrule(lr){8-10} \cmidrule(lr){11-13}
\textbf{Configuration} & P & R & F1 & P & R & F1 & P & R & F1 & P & R & F1 \\
\midrule
Heuristic Baseline       & 2.0 & 66.1 & 4.0 & 1.7 & 73.8 & 3.3 & 1.6 & 59.1 & 3.0 & 1.0 & 47.0 & 2.0 \\
\midrule
Claude Code              & 40.6 & 56.1 & 47.1 & 53.2 & 67.6 & 59.6 & 29.5 & 43.1 & 35.0 & 65.1 & 46.3 & 54.1 \\
Codex CLI                & 43.4 & 57.4 & \textbf{49.4} & 53.3 & 69.4 & 60.3 & 32.7 & 43.6 & \textbf{37.4} & 64.4 & 46.5 & 54.0 \\
OpenCode (Sonnet)        & 44.2 & 53.3 & 48.3 & 52.6 & 66.8 & 58.8 & 33.3 & 38.3 & 35.6 & 62.1 & 43.5 & 51.2 \\
\midrule
OpenCode (Qwen)          & 40.9 & 58.7 & 48.2 & 53.0 & 70.4 & 60.5 & 29.8 & 45.4 & 36.0 & 62.1 & 48.2 & \textbf{54.3} \\
OpenCode (GLM)           & 43.2 & 57.4 & 49.3 & 53.0 & 71.0 & 60.6 & 32.6 & 43.1 & 37.1 & 63.6 & 46.7 & 53.9 \\
OpenCode (Kimi)          & 44.1 & 53.9 & 48.5 & 56.1 & 66.2 & \textbf{60.7} & 32.1 & 40.4 & 35.8 & 67.7 & 43.0 & 52.6 \\
OpenCode (DeepSeek)      & 39.3 & 54.7 & 45.7 & 52.9 & 65.8 & 58.6 & 27.8 & 41.8 & 33.4 & 59.6 & 43.0 & 50.0 \\
\bottomrule
\end{tabular}
}
\end{table}

\begin{table*}[htbp]
\centering
\caption{Update results (\%). \textit{Exec}: executability score; \textit{Cov}: coverage overlap score; \textit{Mod}: modification similarity; \textit{OA}: overall composite score. Best OA per type column is \textbf{bolded}.}
\label{tab:rq2}
\renewcommand{\arraystretch}{1.15}
\resizebox{0.8\linewidth}{!}{
\begin{tabular}{l rrrr rrrr rrrr rrrr}
\toprule
& \multicolumn{4}{c}{\textbf{Overall}} & \multicolumn{4}{c}{\textbf{Breaking}} & \multicolumn{4}{c}{\textbf{Stale}} & \multicolumn{4}{c}{\textbf{Missing}} \\
\cmidrule(lr){2-5} \cmidrule(lr){6-9} \cmidrule(lr){10-13} \cmidrule(lr){14-17}
\textbf{Configuration} & Exec & Cov & Mod & OA & Exec & Cov & Mod & OA & Exec & Cov & Mod & OA & Exec & Cov & Mod & OA \\
\midrule
Claude Code              & 97.0 & 90.2 & 51.0 & 70.5 & 94.8 & 90.7 & 65.0 & 73.2 & 97.3 & 90.7 & 45.8 & 68.5 & 96.7 & 88.5 & 39.1 & 63.8 \\
Codex CLI                & 99.2 & 79.2 & 50.3 & \textbf{72.3} & 98.5 & 71.4 & 63.3 & \textbf{76.6} & 99.3 & 79.0 & 45.8 & \textbf{70.8} & 99.5 & 73.2 & 38.1 & \textbf{65.7} \\
OpenCode (Sonnet)        & 96.2 & 77.3 & 47.6 & 68.9 & 93.6 & 69.9 & 62.5 & 73.8 & 95.9 & 76.8 & 41.0 & 65.4 & 96.0 & 71.8 & 36.9 & 62.6 \\
\midrule
OpenCode (Qwen)          & 94.9 & 83.1 & 48.6 & 67.0 & 94.8 & 83.9 & 64.6 & 73.3 & 94.2 & 82.2 & 41.7 & 62.8 & 93.5 & 80.0 & 37.0 & 59.1 \\
OpenCode (GLM)           & 95.5 & 87.8 & 47.8 & 69.3 & 94.2 & 88.2 & 59.7 & 73.0 & 95.4 & 88.5 & 41.7 & 66.7 & 94.2 & 86.2 & 36.4 & 62.2 \\
OpenCode (Kimi)          & 87.7 & 80.1 & 54.0 & 63.6 & 90.7 & 84.0 & 70.9 & 68.3 & 84.8 & 79.2 & 46.3 & 59.8 & 85.9 & 77.8 & 42.1 & 56.0 \\
OpenCode (DeepSeek)      & 91.4 & 80.4 & 50.3 & 64.5 & 91.0 & 82.6 & 65.4 & 70.4 & 90.3 & 80.7 & 43.6 & 60.9 & 88.9 & 76.5 & 38.4 & 55.4 \\
\bottomrule
\end{tabular}
}
\end{table*}

The seven LLM-based configurations achieve remarkably similar overall F1 scores, ranging from 45.7\% to 49.4\%, with less than four percentage points separating the strongest from the weakest. This convergence holds across closed-source flagships and open-weight backbones, as well as across proprietary and open-source agent frameworks. When the framework is held constant, five backbones evaluated under OpenCode span only 3.6 F1 points; when the backbone is held constant, the Claude Code and OpenCode configurations sharing Claude Sonnet 4.6 differ by 1.2 F1 points. Beneath this aggregate convergence, all seven configurations exhibit a systematic Recall-over-Precision imbalance, with Recall exceeding Precision by between 9.1 and 17.8 percentage points (mean of 13.7). No configuration deviates from this pattern, which indicates a shared inductive bias toward over-prediction rather than an idiosyncratic property of any single backbone. Together, these observations suggest that the performance bottleneck lies not in any specific framework or backbone, but in the inherent difficulty of project-level test identification.

The three evolution types exhibit substantially different difficulty, and the relative ordering is preserved across all seven configurations. Test-Breaking is the easiest, with an average F1 of 59.9\% and a tight spread of 2.1 points across configurations, since explicit execution failure signals are available to locate affected tests. Test-Missing occupies an intermediate position with an average F1 of 52.9\%, where relatively high Precision is paired with lower Recall, indicating that configurations recognize some scenarios requiring new tests but miss over half of them. Test-Stale is by far the hardest, with an average F1 of 35.8\% and both Precision and Recall substantially depressed. Because stale tests still pass on the updated code, no execution signal indicates that updates are needed, and configurations must rely entirely on proactive semantic reasoning, a capability that the seven evaluated systems lack in roughly equal measure.

The heuristic baseline provides an informative reference point. Its Recall of 66.1\% surpasses every LLM-based configuration, while its Precision is only 2.0\% with over 39{,}000 false positives. The contrast is particularly striking on Test-Stale, where the heuristic's Recall of 59.1\% exceeds the seven-configuration average of 42.2\% by 16.9 percentage points. Even this exhaustive one-hop dependency analysis fails to reach 100\% Recall on any type, with approximately one-third of truly affected tests remaining undetected, which indicates that a substantial portion of test-code dependencies operate through indirect channels such as multi-hop call chains, shared state, or implicit semantic coupling that structural analysis cannot capture.

\finding{1}{
Across seven configurations spanning three agent frameworks and six base models, identification F1 remains within 45.7\% to 49.4\%, with backbone variation contributing only 3.6 F1 points and framework variation only 1.2 points. All configurations exhibit a systematic Recall-over-Precision imbalance, with the gap ranging from 9.1 to 17.8 percentage points, which reveals a shared bias toward over-prediction. Test-Stale is the hardest type, with an average F1 of approximately 36\%, because it requires proactive semantic reasoning without execution signals, and even exhaustive structural analysis misses about one-third of affected tests.
}

\subsection{RQ2: Update Effectiveness}
\label{sec:results:rq2}

Table~\ref{tab:rq2} presents the update quality metrics for the seven LLM-based configurations. Overall composite scores cluster within a band of 8.8 percentage points, ranging from 63.6\% to 72.3\%, with Codex CLI achieving the highest score on every type column.

Configurations achieve high executability scores, ranging from 87.7\% to 99.2\%, indicating that producing compilable and runnable test modifications is largely tractable. Coverage overlap differs more substantially across configurations, with Claude Code attaining 90.2\% and OpenCode with the GLM backbone attaining 87.8\%, both well above the 77\% to 84\% range of the remaining five configurations. This advantage in coverage does not translate directly into higher composite scores, as Codex CLI leads at 72.3\% despite a coverage overlap of only 79.2\%, while Claude Code lags at 70.5\% with the highest coverage. The pattern reflects a recurring dimensional trade-off: Claude Code applies more aggressive modifications that improve coverage at the cost of marginally lower executability of 97.0\% relative to Codex CLI's 99.2\%, whereas OpenCode with the Kimi backbone pursues higher modification fidelity at 54.0\%, the highest across configurations, at the cost of pronounced executability degradation to 87.7\%, the lowest across configurations, which depresses its composite score to 63.6\%.

The modification similarity score is the lowest sub-metric across all configurations and evolution types, ranging from 36.4\% to 70.9\%, and falls 33.7 to 48.9 percentage points below the corresponding executability score within each configuration. This systematic gap indicates that current configurations can produce executable test modifications whose surface form diverges substantially from how developers actually update tests. The gap widens further on Test-Missing, where modification similarity drops to between 36.4\% and 42.1\%, reflecting the inherently larger implementation space when generating new tests rather than revising existing assertions. The pattern argues against treating executability as a sufficient proxy for update quality, since high executability can mask substantial divergence from developer intent.

The type-wise difficulty ranking on the update task differs from that on identification. Update difficulty follows the order Test-Breaking with an average composite score of 72.7\%, Test-Stale at 65.0\%, and Test-Missing at 60.7\%, whereas identification difficulty ranks Test-Breaking, Test-Missing, and Test-Stale. This crossover indicates that Test-Stale is hardest to identify but not hardest to update, since stale tests require only targeted assertion changes once located, while Test-Missing is easier to identify but harder to update because it demands generating entirely new code that naturally produces lower similarity to GT and lower coverage overlap.

\finding{2}{
The seven configurations cluster within an 8.8-point band on the composite update score, ranging from 63.6\% to 72.3\%. Executability remains consistently high (87.7\% to 99.2\%) yet exceeds modification similarity by 33.7 to 48.9 percentage points within each configuration, indicating that producing executable tests is far easier than producing tests aligned with developer intent. The type-wise difficulty ranking flips between the two subtasks: Test-Stale is hardest to identify but not to update, whereas Test-Missing exhibits the inverse pattern.
}

\subsection{RQ3: Impact of Task Characteristics}
\label{sec:results:rq3}

To understand what makes test evolution tasks difficult, we analyze configuration performance, averaged across the seven LLM-based configurations, along three task characteristic dimensions: evolution type composition, source change scale, and test change scope. Table~\ref{tab:rq3} summarizes the results.

\begin{table}[htbp]
\centering
\caption{Impact of task characteristics on configuration performance, averaged across the seven LLM-based configurations. \textit{Id-F1}: identification F1 (\%); \textit{Up-OA}: update overall score (\%).}
\label{tab:rq3}
\renewcommand{\arraystretch}{1.15}
\resizebox{0.9\linewidth}{!}{
\begin{tabular}{llrrr}
\toprule
\textbf{Dimension} & \textbf{Group} & \textbf{N} & \textbf{Id-F1} & \textbf{Up-OA} \\
\midrule
\multirow{7}{*}{\shortstack[l]{Type\\Composition}}
& Breaking-only                  &  58 & 62.0 & 84.2 \\
& Stale-only                     &  33 & 33.1 & 78.4 \\
& Missing-only                   &   4 & 68.8 & 41.7 \\
\cmidrule(lr){2-5}
& Breaking + Missing             &  45 & 74.3 & 63.5 \\
& Breaking + Stale + Missing     &  45 & 64.8 & 65.5 \\
& Breaking + Stale               &  24 & 29.8 & 75.4 \\
& Stale + Missing                & 105 & 34.8 & 58.2 \\
\midrule
\multirow{3}{*}{\shortstack[l]{Source Change\\Scale}}
& Small ($\leq$19 lines)         & 102 & 46.0 & 71.4 \\
& Medium (20--55 lines)          & 100 & 52.1 & 68.4 \\
& Large ($\geq$56 lines)         & 112 & 46.1 & 64.6 \\
\midrule
\multirow{3}{*}{\shortstack[l]{Test Change\\Scope}}
& Small (1 method)               &  81 & 22.7 & 61.9 \\
& Medium (2--3 methods)          & 102 & 46.6 & 70.4 \\
& Large ($\geq$4 methods)        & 131 & 53.2 & 69.9 \\
\bottomrule
\end{tabular}
}
\end{table}

The type composition dimension reveals how the co-occurrence of different evolution types shapes task difficulty. Single-type tasks serve as instructive baselines: Breaking-only tasks achieve the highest update score of 84.2\% with solid identification F1 of 62.0\%, while Stale-only tasks are substantially harder to identify with F1 of 33.1\% but still yield high update scores of 78.4\% once the correct tests are located, which confirms that stale tests are difficult to find but require only targeted modifications. Missing-only tasks are too few in number to support robust conclusions, although their low update score of 41.7\% hints at the difficulty of generating new tests from scratch. Among multi-type combinations, Breaking + Missing achieves the highest identification F1 of 74.3\%, since breaking tests provide explicit failure signals that anchor the search process. Once Test-Stale enters the combination, identification F1 drops sharply regardless of what other types are present, with Breaking + Stale falling to 29.8\% and Stale + Missing falling to 34.8\%. This pattern suggests that Test-Stale acts as a ``poison factor'' in identification, since the absence of any execution signal undermines the systematic location of all affected tests. A noteworthy exception arises when all three types co-occur: Breaking + Stale + Missing recovers to an identification F1 of 64.8\%, well above Breaking + Stale alone, suggesting that the explicit signals from Missing components partially compensate for the disorientation introduced by Stale. An interesting contrast emerges in the update dimension, where Breaking + Stale tasks achieve the highest multi-type update score of 75.4\% despite having among the lowest identification F1, indicating that once located, the required updates for breaking and stale types are relatively straightforward compared to generating missing tests.

The source change scale dimension reveals a non-monotonic pattern in identification performance. Medium-scale changes between 20 and 55 lines yield the highest identification F1 of 52.1\%, while both small-scale changes at 46.0\% and large-scale changes at 46.1\% are more difficult. Small source diffs provide insufficient contextual information for configurations to infer the scope of test impact, while large diffs present excessive information that complicates focusing on the most relevant changes. Update quality, in contrast, decreases monotonically from 71.4\% for small changes to 64.6\% for large changes, consistent with the intuition that larger source changes require more extensive and complex test modifications.

The test change scope dimension reveals a counterintuitive pattern in which identification F1 increases from 22.7\% for single-method tasks to 53.2\% for tasks involving four or more methods. This is not because single-method tasks are inherently harder to understand. Rather, configurations exhibit a consistent tendency to modify a similar number of test methods regardless of task size: on single-method tasks, configurations predict approximately 3.6 methods on average, well above the affected count, producing massive over-prediction that collapses Precision to 13.6\%. On large tasks, the prediction volume aligns more naturally with the GT, yielding higher Precision of 53.2\%. This finding points to a fundamental limitation, namely that current configurations lack the ability to calibrate their modification scope to the actual task requirements, applying a roughly constant effort budget regardless of whether the task demands touching one method or ten.

\finding{3}{
Three factors shape task difficulty. First, Test-Stale acts as a poison factor that sharply reduces identification performance in mixed-type tasks, although the presence of explicit Missing signals partially compensates for this effect. Second, medium-scale source changes are easier to identify than both small and large ones, forming an inverted-U pattern. Third, configurations lack scope calibration, causing severe over-prediction on small tasks.
}

\subsection{RQ4: Failure Analysis}
\label{sec:results:rq4}

\begin{figure*}[htbp]
  \centering
  \includegraphics[width=\linewidth]{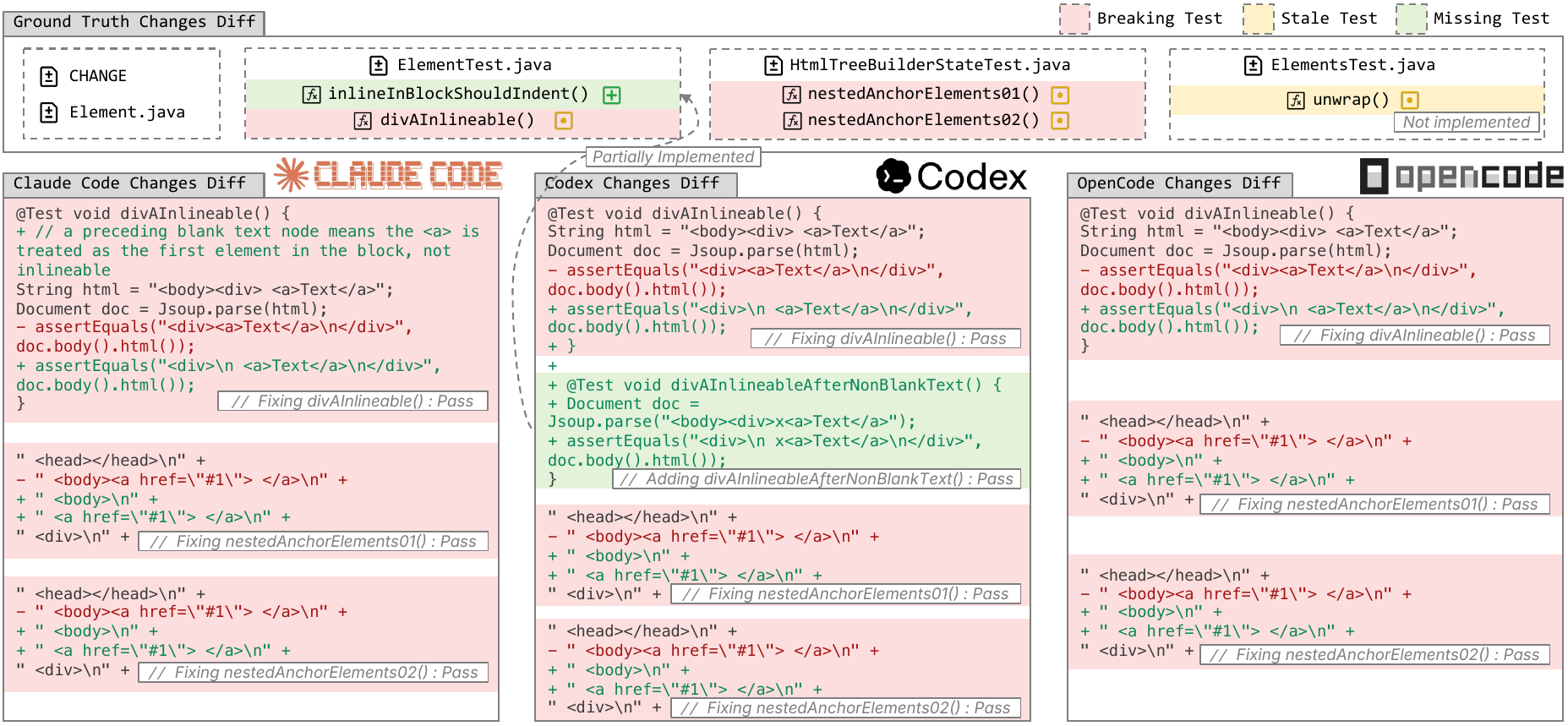}
  \caption{Case study on Task 293 (jsoup): GT changes (top) versus actual modifications produced by each configuration (bottom).}
  \label{fig:rq4_diff}
\end{figure*}

We select the jsoup motivating example from Section~\ref{sec:motivation} for in-depth analysis, as it is the most representative task in TEBench: it simultaneously involves all three evolution types, requires updates to five test methods across three files, and exhibits the ``small change, wide impact'' characteristic, with only 12 source-code lines changed. Figure~\ref{fig:rq4_diff} compares the GT modifications with those produced by the three industrial agent frameworks (Claude Code, Codex CLI, and OpenCode under the Sonnet backbone), which serve as a representative cross-section of the seven configurations evaluated in TEBench.

All three depicted configurations successfully fix the three Breaking tests, correctly updating the expected HTML strings. Claude Code additionally adds a semantic comment to \texttt{divAInlineable}, explaining the root cause of the change. However, none of these configurations updates the Stale test \texttt{unwrap}. The four OpenCode configurations using open-weight backbones exhibit qualitatively similar patterns, none of which extends materially beyond the cross-section depicted in Figure~\ref{fig:rq4_diff}.

For Test-Missing, Codex CLI partially addresses the missing coverage by generating a new test. This test is not entirely misaligned with the GT, since it captures one representative scenario also covered by \texttt{inlineInBlockShouldIndent}, namely the case where an inline element follows non-blank preceding text. However, the GT test is broader and behavior-oriented, systematically verifying consistent indentation across three distinct input variants, including non-blank text, a newline, and a blank space before the inline element. In contrast, Codex CLI covers only one of these scenarios, leaving the broader behavioral consistency unchecked. This suggests that Codex CLI identifies part of the newly exposed behavior but does not recover the full semantic scope reflected in the developer-written test.

Since Codex CLI is the only configuration that both fixes all Breaking tests and attempts to generate a new test, we examine its execution trajectory to understand the underlying problem-solving strategy. Codex CLI begins by inspecting the source diff and searching for related test keywords, then runs \texttt{ElementTest} in isolation, where only \texttt{divAInlineable} fails. It fixes this assertion and adds its new test, but the new test's assertion is initially incorrect, requiring an additional fix-rerun cycle. Only when Codex CLI later runs the full test suite does it discover the two failures in \texttt{HtmlTreeBuilderStateTest}, which resides in a different package. After patching these, all tests pass with adequate JaCoCo coverage, and Codex CLI terminates. This trajectory exposes two fundamental limitations. First, the configuration's identification strategy is entirely execution-driven, since it discovers affected tests through test failures rather than through proactive reasoning about change impact: the \texttt{HtmlTreeBuilderStateTest} failures were found by running the full suite rather than by analyzing cross-package dependencies. Second, the configuration's termination is governed by the joint condition of all tests passing and adequate coverage being met, which structurally prevents it from detecting stale tests. The \texttt{unwrap} test passes because \texttt{stripNewlines()} masks the formatting change, and no execution signal alerts the configuration to this semantic gap.

\finding{4}{
Current configurations operate in a reactive ``execute-fail-fix'' loop that succeeds for Test-Breaking but cannot address Test-Stale, since no failure signal is available, nor can they fully cover Test-Missing, since the loop encourages branch-level patching rather than behavior-level test design.
}

\section{Related Work}
\label{sec:related}
 
\noindent\textbf{Test Generation and Test Evolution.}
Automated test generation has been extensively studied. Traditional search-based tools such as EvoSuite~\cite{fraser2011evosuite} and constraint-based approaches~\cite{lukasczyk2022pynguin} maximize structural coverage through evolutionary algorithms. With the rise of LLMs, recent work has shifted toward more natural test generation: ChatUniTest~\cite{chen2024chatunitest} introduced a generation-validation-repair loop, HITS~\cite{wang2024hits} decomposes methods via slicing for incremental generation, and CoverUp~\cite{altmayerpizzorno2025coverup} iteratively targets uncovered lines. Other notable approaches further improve coverage and quality through program analysis, execution path guidance, and multi-agent collaboration~\cite{lemieux2023codamosa, schaefer2023empirical, ryan2024code, yang2024enhancing, yuan2024evaluating, pan2025aster, wang2024testagent, jain2025testgeneval, wang2025testeval}.
 
However, as LLM-based coding agents become increasingly embedded in development workflows, test \emph{evolution}---updating existing tests in response to code changes---emerges as a more common and practical need than generating tests from scratch. Researchers have studied this challenge from both empirical~\cite{zaidman2011studying, marsavina2014studying} and automation perspectives. SITAR~\cite{wang2021understanding} identified factors influencing test updates through a large-scale empirical study. CEPROT~\cite{hu2023identify} proposed a Transformer-based approach for identifying and updating obsolete tests given method-level code changes. REACCEPT~\cite{chi2025reaccept} integrated LLMs with dynamic validation to automate both identification and updating. Other recent work has further advanced test repair and update techniques~\cite{sun2023revisiting, liu2024fix, yaraghi2025automated, zhang2025unit}. Several benchmarks have been proposed alongside these methods, such as Updates4J~\cite{zhang2025unit} with 195 samples, but they all adopt a \emph{method-level paired} formulation where the mapping between changed code and affected tests is pre-given. TEBench is the first to define a project-level task requiring the system to autonomously identify tests requiring modification and determine where new tests are needed from the full project context.
 
\noindent\textbf{Coding Agents and SE Benchmarks.}
LLM-based coding agents such as SWE-agent~\cite{yang2024swe}, OpenHands~\cite{wang2025openhands}, Claude Code~\cite{anthropic2025claude}, Codex CLI~\cite{openai2025codex}, and OpenCode~\cite{opencodecontributors2025opencode} have demonstrated strong capabilities on code understanding and repair~\cite{xia2025demystifying}, yet none have been systematically evaluated on test evolution tasks. On the benchmarking side, SWE-bench~\cite{jimenez2024swe} established the paradigm of repository-level issue resolution, inspiring extensions toward long-horizon evolution and enterprise complexity~\cite{chowdhury2024swebenchverified, thai2025swe, deng2025swe}. SWT-bench~\cite{muendler2024swt} shifted focus to test generation for known bugs~\cite{ahmed2024tdd}. TEBench targets a complementary dimension: evolving existing test suites alongside production code changes, bridging the gap between code repair benchmarks and test generation benchmarks.

\section{Threats to Validity}
\label{sec:threats}
 
\noindent\textbf{Internal Validity.}
The GT is derived from developer-written test modifications in real commits, which constitutes a faithful record of actual developer intent rather than an arbitrary gold standard, even though it may not represent the unique correct solution. Functionally equivalent updates with different implementations could be penalized by our modification similarity metric. To mitigate this, we employ multiple complementary metrics: executability is entirely objective; coverage overlap remains meaningful regardless of implementation differences; and modification similarity is interpreted alongside the other dimensions rather than in isolation, ensuring that semantically valid alternatives are not disproportionately penalized.

\noindent\textbf{External Validity.}
TEBench covers 10 Java projects from the Defects4J ecosystem, which may limit generalizability to other languages or project types. The uneven distribution across projects reflects natural variation in test maintenance activity rather than sampling bias. We mitigate this by reporting per-project results alongside aggregate metrics, and note that extending TEBench to additional languages and projects is a natural direction for future work.

\noindent\textbf{Construct Validity.}
Our evaluation measures coverage overlap with the GT rather than absolute coverage improvement, based on the rationale that test evolution should align with developer intent for the specific change rather than maximize coverage indiscriminately. Our unified task prompt was designed for fairness across all configurations; configuration-specific prompt tuning could yield higher absolute performance but would introduce confounds that undermine cross-configuration comparability. Our setup therefore prioritizes standardization and reproducibility over per-configuration optimization.

\section{Conclusion}
\label{sec:conclusion}
 
We presented TEBench, the first project-level benchmark for test evolution, which requires systems to autonomously identify tests requiring modification and determine where new tests are needed given a project repository and a code-changing commit. TEBench comprises 314 task instances from 10 real-world Java projects, covering three evolution types: Test-Breaking, Test-Stale, and Test-Missing. Our evaluation of seven configurations spanning three industrial agent frameworks and six base models reveals that current systems achieve an identification F1 of only 45.7\% to 49.4\%, with Test-Stale posing the greatest challenge at an average F1 of approximately 36\% due to the absence of execution failure signals, which fundamentally limits the ability of current systems to detect semantically outdated tests or proactively generate missing ones.

These findings point to several directions for future work. First, integrating static dependency analysis with LLM-based semantic reasoning could combine the high recall of structural approaches with the precision of language understanding. Second, developing systems that reason about testing intent beyond execution signals, for instance by analyzing coverage gaps, inferring behavioral contracts from code changes, or aligning modifications more closely with developer intent, could address both the Test-Stale bottleneck and the systematic divergence between executable and developer-aligned test updates. Third, extending TEBench to additional programming languages and larger-scale industrial projects would further validate the generalizability of our findings. We hope that TEBench serves as a catalyst for advancing test evolution capabilities in coding agents.

\bibliography{sample-base}

@InProceedings{fraser2011evosuite,
  author    = {Fraser, Gordon and Arcuri, Andrea},
  booktitle = {Proceedings of the 19th ACM SIGSOFT Symposium and the 13th European Conference on Foundations of Software Engineering (ESEC/FSE)},
  title     = {{EvoSuite}: Automatic Test Suite Generation for Object-Oriented Software},
  year      = {2011},
  pages     = {416--419},
}

@InProceedings{lukasczyk2022pynguin,
  author    = {Lukasczyk, Stephan and Fraser, Gordon},
  booktitle = {Proceedings of the 44th IEEE/ACM International Conference on Software Engineering: Companion Proceedings (ICSE-Companion)},
  title     = {{Pynguin}: Automated Unit Test Generation for {Python}},
  year      = {2022},
  pages     = {168--172},
}

@InProceedings{lemieux2023codamosa,
  author    = {Lemieux, Caroline and Inala, Jeevana Priya and Lahiri, Shuvendu K and Sen, Siddhartha},
  booktitle = {Proceedings of the 45th IEEE/ACM International Conference on Software Engineering (ICSE)},
  title     = {{CodaMOSA}: Escaping Coverage Plateaus in Test Generation with Pre-Trained Large Language Models},
  year      = {2023},
  pages     = {919--931},
}

@Article{schaefer2023empirical,
  author  = {Sch{\"a}fer, Max and Nadi, Sarah and Eghbali, Aryaz and Tip, Frank},
  journal = {IEEE Transactions on Software Engineering (TSE)},
  title   = {An Empirical Evaluation of Using Large Language Models for Automated Unit Test Generation},
  year    = {2023},
  volume  = {50},
  number  = {1},
  pages   = {85--105},
}

@InProceedings{chen2024chatunitest,
  author    = {Chen, Yinghao and Hu, Zehao and Zhi, Chen and Han, Junxiao and Deng, Shuiguang and Yin, Jianwei},
  booktitle = {Companion Proceedings of the 32nd ACM International Conference on the Foundations of Software Engineering (FSE-Companion)},
  title     = {{ChatUniTest}: A Framework for {LLM}-Based Test Generation},
  year      = {2024},
  pages     = {572--576},
}

@InProceedings{wang2024hits,
  author    = {Wang, Zejun and Liu, Kaibo and Li, Ge and Jin, Zhi},
  booktitle = {Proceedings of the 39th IEEE/ACM International Conference on Automated Software Engineering (ASE)},
  title     = {{HITS}: High-Coverage {LLM}-Based Unit Test Generation via Method Slicing},
  year      = {2024},
  pages     = {1258--1268},
}

@Article{altmayerpizzorno2025coverup,
  author  = {Altmayer Pizzorno, Juan and Berger, Emery D},
  journal = {Proceedings of the ACM on Software Engineering (PACMSE)},
  title   = {{CoverUp}: Effective High Coverage Test Generation for {Python}},
  year    = {2025},
  volume  = {2},
  number  = {FSE},
  pages   = {2897--2919},
}

@Article{ryan2024code,
  author  = {Ryan, Gabriel and Jain, Siddhartha and Shang, Mingyue and Wang, Shiqi and Ma, Xiaofei and Ramanathan, Murali Krishna and Ray, Baishakhi},
  journal = {Proceedings of the ACM on Software Engineering (PACMSE)},
  title   = {Code-Aware Prompting: A Study of Coverage-Guided Test Generation in Regression Setting Using {LLM}},
  year    = {2024},
  volume  = {1},
  number  = {FSE},
  pages   = {951--971},
}

@Article{yuan2024evaluating,
  author  = {Yuan, Zhiqiang and Liu, Mingwei and Ding, Shiji and Wang, Kaixin and Chen, Yixuan and Peng, Xin and Lou, Yiling},
  journal = {Proceedings of the ACM on Software Engineering (PACMSE)},
  title   = {Evaluating and Improving {ChatGPT} for Unit Test Generation},
  year    = {2024},
  volume  = {1},
  number  = {FSE},
  pages   = {1703--1726},
}

@Article{yang2024enhancing,
  author  = {Yang, Chen and Chen, Junjie and Lin, Bin and Zhou, Jianyi and Wang, Ziqi},
  journal = {arXiv preprint arXiv:2404.04966},
  title   = {Enhancing {LLM}-Based Test Generation for Hard-to-Cover Branches via Program Analysis},
  year    = {2024},
}

@InProceedings{pan2025aster,
  author    = {Pan, Rangeet and Kim, Myeongsoo and Krishna, Rahul and Pavuluri, Raju and Sinha, Saurabh},
  booktitle = {Proceedings of the 47th IEEE/ACM International Conference on Software Engineering: Software Engineering in Practice (ICSE-SEIP)},
  title     = {{ASTER}: Natural and Multi-Language Unit Test Generation with {LLMs}},
  year      = {2025},
  pages     = {413--424},
}

@Article{wang2024testagent,
  author  = {Wang, Zhijie and Liu, Mingqi and Chu, Zhaoyang and Wang, Wenhan and Song, Da and Ma, Lei},
  journal = {arXiv preprint arXiv:2401.01602},
  title   = {{TestAgent}: An {LLM}-Based Multi-Agent System for Automated Unit Test Generation},
  year    = {2024},
}

@InProceedings{jain2025testgeneval,
  author    = {Jain, Kush and Synnaeve, Gabriel and Rozi{\`{e}}re, Baptiste},
  booktitle = {The Thirteenth International Conference on Learning Representations (ICLR)},
  title     = {{TestGenEval}: A Real World Unit Test Generation and Test Completion Benchmark},
  year      = {2025},
}

@InProceedings{wang2025testeval,
  author    = {Wang, Wenhan and Yang, Chenyuan and Wang, Zhijie and Huang, Yuheng and Chu, Zhaoyang and Song, Da and Zhang, Lingming and Chen, An Ran and Ma, Lei},
  booktitle = {Findings of the Association for Computational Linguistics: NAACL 2025},
  title     = {{TestEval}: Benchmarking Large Language Models for Test Case Generation},
  year      = {2025},
  pages     = {3547--3562},
}

@Article{zaidman2011studying,
  author  = {Zaidman, Andy and Van Rompaey, Bart and Van Deursen, Arie and Demeyer, Serge},
  journal = {Empirical Software Engineering},
  title   = {Studying the Co-Evolution of Production and Test Code in Open Source and Industrial Developer Test Processes through Repository Mining},
  year    = {2011},
  volume  = {16},
  number  = {3},
  pages   = {325--364},
}

@InProceedings{marsavina2014studying,
  author    = {Marsavina, Cosmin and Romano, Daniele and Zaidman, Andy},
  booktitle = {Proceedings of the 14th IEEE International Working Conference on Source Code Analysis and Manipulation (SCAM)},
  title     = {Studying Fine-Grained Co-Evolution Patterns of Production and Test Code},
  year      = {2014},
  pages     = {195--204},
}

@InProceedings{wang2021understanding,
  author    = {Wang, Sinan and Wen, Ming and Liu, Yepang and Wang, Ying and Wu, Rongxin},
  booktitle = {Proceedings of the 28th IEEE International Conference on Software Analysis, Evolution and Reengineering (SANER)},
  title     = {Understanding and Facilitating the Co-Evolution of Production and Test Code},
  year      = {2021},
  pages     = {272--283},
}

@InProceedings{hu2023identify,
  author    = {Hu, Xing and Liu, Zhuang and Xia, Xin and Liu, Zhongxin and Xu, Tongtong and Yang, Xiaohu},
  booktitle = {Proceedings of the 38th IEEE/ACM International Conference on Automated Software Engineering (ASE)},
  title     = {Identify and Update Test Cases When Production Code Changes: A Transformer-Based Approach},
  year      = {2023},
  pages     = {1111--1122},
}

@Article{sun2023revisiting,
  author  = {Sun, Weifeng and Yan, Meng and Liu, Zhongxin and Xia, Xin and Lei, Yan and Lo, David},
  journal = {ACM Transactions on Software Engineering and Methodology (TOSEM)},
  title   = {Revisiting the Identification of the Co-Evolution of Production and Test Code},
  year    = {2023},
  volume  = {32},
  number  = {6},
  pages   = {1--37},
}

@InProceedings{liu2024fix,
  author    = {Liu, Jun and Yan, Jiwei and Xie, Yuanyuan and Yan, Jun and Zhang, Jian},
  booktitle = {Proceedings of the 35th IEEE International Symposium on Software Reliability Engineering (ISSRE)},
  title     = {Fix the Tests: Augmenting {LLMs} to Repair Test Cases with Static Collector and Neural Reranker},
  year      = {2024},
  pages     = {367--378},
}

@Article{yaraghi2025automated,
  author  = {Yaraghi, Ahmadreza Saboor and Holden, Darren and Kahani, Nafiseh and Briand, Lionel C.},
  journal = {IEEE Transactions on Software Engineering (TSE)},
  title   = {Automated Test Case Repair Using Language Models},
  year    = {2025},
  volume  = {51},
  number  = {4},
  pages   = {1104--1133},
}

@Article{chi2025reaccept,
  author  = {Chi, Jianlei and Wang, Xiaotian and Huang, Yuhan and Yu, Lechen and Cui, Di and Sun, Jianguo and Sun, Jun},
  journal = {Proceedings of the ACM on Software Engineering (PACMSE)},
  title   = {{REACCEPT}: Automated Co-Evolution of Production and Test Code Based on Dynamic Validation and Large Language Models},
  year    = {2025},
  volume  = {2},
  number  = {ISSTA},
  pages   = {1234--1256},
}

@Article{zhang2025unit,
  author  = {Zhang, Yuanhe and Yang, Zhiquan and Pan, Shengyi and Liu, Zhongxin},
  journal = {arXiv preprint arXiv:2509.24419},
  title   = {Unit Test Update through {LLM}-Driven Context Collection and Error-Type-Aware Refinement},
  year    = {2025},
}

@Article{rahman2025utfix,
  author  = {Rahman, Shanto and Kuhar, Sachit and Cirisci, Berk and Garg, Pranav and Wang, Shiqi and Ma, Xiaofei and Deoras, Anoop and Ray, Baishakhi},
  journal = {Proceedings of the ACM on Programming Languages (PACMPL)},
  title   = {{UTFix}: Change Aware Unit Test Repairing Using {LLM}},
  year    = {2025},
  volume  = {9},
  number  = {OOPSLA1},
  pages   = {143--168},
}

@InProceedings{just2014defects4j,
  author    = {Just, Ren{\'e} and Jalali, Darioush and Ernst, Michael D},
  booktitle = {Proceedings of the 2014 International Symposium on Software Testing and Analysis (ISSTA)},
  title     = {{Defects4J}: A Database of Existing Faults to Enable Controlled Testing Studies for {Java} Programs},
  year      = {2014},
  pages     = {437--440},
}

@InProceedings{jimenez2024swe,
  author    = {Jimenez, Carlos E and Yang, John and Wettig, Alexander and Yao, Shunyu and Pei, Kexin and Press, Ofir and Narasimhan, Karthik R},
  booktitle = {The Twelfth International Conference on Learning Representations (ICLR)},
  title     = {{SWE-Bench}: Can Language Models Resolve Real-World {GitHub} Issues?},
  year      = {2024},
  url       = {https://openreview.net/forum?id=VTF8yNQM66},
}

@Article{chowdhury2024swebenchverified,
  author  = {Chowdhury, Neil and Aider, James and Cassano, Federico and Zhuo, Jiawei and Liu, Qian and Jimenez, Carlos E. and Narasimhan, Karthik and Press, Ofir},
  journal = {arXiv preprint arXiv:2406.12952},
  title   = {{SWE-Bench} Verified: A Stricter Evaluation for {AI} Software Engineering},
  year    = {2024},
}

@Article{deng2025swe,
  author  = {Deng, Xiang and Da, Jeff and Pan, Edwin and He, Yannis Yiming and Ide, Charles and Garg, Kanak and Lauffer, Niklas and Park, Andrew and Pasari, Nitin and Rane, Chetan and others},
  journal = {arXiv preprint arXiv:2509.16941},
  title   = {{SWE-Bench} Pro: Can {AI} Agents Solve Long-Horizon Software Engineering Tasks?},
  year    = {2025},
}

@Article{thai2025swe,
  author  = {Thai, Minh VT and Le, Tue and Manh, Dung Nguyen and Nhat, Huy Phan and Bui, Nghi DQ},
  journal = {arXiv preprint arXiv:2512.18470},
  title   = {{SWE-Evo}: Benchmarking Coding Agents in Long-Horizon Software Evolution Scenarios},
  year    = {2025},
}

@Article{muendler2024swt,
  author  = {M{\"u}ndler, Niels and M{\"u}ller, Mark N and He, Jingxuan and Vechev, Martin},
  journal = {Advances in Neural Information Processing Systems (NeurIPS)},
  title   = {{SWT-Bench}: Testing and Validating Real-World Bug-Fixes with Code Agents},
  year    = {2024},
  volume  = {37},
  pages   = {81857--81887},
}

@Article{ahmed2024tdd,
  author  = {Ahmed, Toufique and Hirzel, Martin and Pan, Rangeet and Shinnar, Avraham and Sinha, Saurabh},
  journal = {arXiv preprint arXiv:2412.02883},
  title   = {{TDD-Bench} Verified: Can {LLMs} Generate Tests for Issues before They Get Resolved?},
  year    = {2024},
}

@Article{yang2024swe,
  author  = {Yang, John and Jimenez, Carlos E and Wettig, Alexander and Lieret, Kilian and Yao, Shunyu and Narasimhan, Karthik and Press, Ofir},
  journal = {Advances in Neural Information Processing Systems (NeurIPS)},
  title   = {{SWE-Agent}: Agent-Computer Interfaces Enable Automated Software Engineering},
  year    = {2024},
  volume  = {37},
  pages   = {50528--50652},
}

@InProceedings{wang2025openhands,
  author    = {Wang, Xingyao and Li, Boxuan and Song, Yufan and Xu, Frank F and Tang, Xiangru and Zhuge, Mingchen and Pan, Jiayi and Song, Yueqi and Li, Bowen and Singh, Jaskirat and others},
  booktitle = {The Thirteenth International Conference on Learning Representations (ICLR)},
  title     = {{OpenHands}: An Open Platform for {AI} Software Developers as Generalist Agents},
  year      = {2025},
}

@Article{xia2025demystifying,
  author  = {Xia, Chunqiu Steven and Deng, Yinlin and Dunn, Soren and Zhang, Lingming},
  journal = {Proceedings of the ACM on Software Engineering (PACMSE)},
  title   = {Demystifying {LLM}-Based Software Engineering Agents},
  year    = {2025},
  volume  = {2},
  number  = {FSE},
  pages   = {801--824},
}

@Misc{anthropic2025claude,
  author       = {{Anthropic}},
  title        = {Claude Code},
  year         = {2025},
  url          = {https://docs.anthropic.com/en/docs/claude-code},
  howpublished = {\url{https://docs.anthropic.com/en/docs/claude-code}},
  note         = {Accessed: June 2025},
}

@Misc{openai2025codex,
  author       = {{OpenAI}},
  title        = {Codex {CLI}},
  year         = {2025},
  url          = {https://github.com/openai/codex},
  howpublished = {\url{https://github.com/openai/codex}},
  note         = {Accessed: June 2025},
}

@Misc{opencodecontributors2025opencode,
  author       = {{OpenCode Contributors}},
  title        = {{OpenCode}: An Open-Source Coding Agent},
  year         = {2025},
  url          = {https://github.com/opencode-ai/opencode},
  howpublished = {\url{https://github.com/opencode-ai/opencode}},
  note         = {Accessed: June 2025},
}

@article{qwen,
  author       = {Qwen Team},
  title        = {Qwen3 Technical Report},
  journal      = {CoRR},
  volume       = {abs/2505.09388},
  year         = {2025}
}

@article{deepseek,
  author       = {DeepSeek{-}AI},
  title        = {DeepSeek-V3.2: Pushing the Frontier of Open Large Language Models},
  journal      = {CoRR},
  volume       = {abs/2512.02556},
  year         = {2025}
}

@article{glm,
  author       = {GLM},
  title        = {{GLM-5:} from Vibe Coding to Agentic Engineering},
  journal      = {CoRR},
  volume       = {abs/2602.15763},
  year         = {2026}
}

@article{kimi,
  author       = {Kimi Team},
  title        = {Kimi {K2.5:} Visual Agentic Intelligence},
  journal      = {CoRR},
  volume       = {abs/2602.02276},
  year         = {2026}
}
\bibliographystyle{icml2026}

\end{document}